\documentclass{PoS}
\def\mcg {MCG~$+$8$-$11$-$11}
\usepackage{epsfig}

\title{Compton processes in the bright AGN MCG~+8-11-11}

\ShortTitle{Compton processes in the bright AGN \mcg}

\author{\speaker{Simona Soldi}\\
        Laboratoire AIM - CNRS - CEA/DSM - Universit\'e Paris Diderot (UMR 7158), CEA Saclay, DSM/IRFU/SAp, F91191 Gif-sur-Yvette, France\\
        E-mail: \email{simona.soldi@cea.fr}
	}
\author{Volker Beckmann\\
	Fran\c{c}ois Arago Centre, APC, Universit\'e Paris Diderot, CNRS/IN2P3, CEA/DSM, Observatoire de Paris, 13 rue Watt, 75205 Paris Cedex 13, France
	}
\author{Neil Gehrels\\
	NASA Goddard Space Flight Center, Astrophysics Science Division, Code 661, Greenbelt, MD 20771, USA
	}
\author{Piotr Lubi\'nski\\	        
        Nicolaus Copernicus Astronomical Center, Polish Academy of Sciences ul. Rabianska 8, 87--100 Torun, Poland 
	}
\author{Claudio Ricci\\	        
        ISDC Data Centre for Astrophysics, Observatoire de Gen\`eve, Chemin d'Ecogia 16, Versoix, Switzerland
	}
\author{Roland Walter\\	        
        ISDC Data Centre for Astrophysics, Observatoire de Gen\`eve, Chemin d'Ecogia 16, Versoix, Switzerland
	}

\abstract{We present preliminary results on the hard X-ray emission properties of the Seyfert 1.5 galaxy \mcg,
	  as observed by \textit{INTEGRAL} and \textit{SWIFT}.
	  All the \textit{INTEGRAL} IBIS/ISGRI data available up to October 2009 have been analysed together with 
	  two \textit{SWIFT}/XRT snapshot observations performed in August and October 2009, quasi-simultaneously 
	  to \textit{INTEGRAL} pointed observations of \mcg.
	  No correlation is observed between the hard X-ray flux and the spectral slope, while the position of the 
	  high-energy cut-off is found to have varied during the \textit{INTEGRAL} observations. This points to a change 
	  in the temperature of the Comptonising medium from a minimum value of $kT_{\rm e} = 30-50 \rm \, keV$ to values larger
	  than 100--150~keV.
	  There is no significant detection of Compton reflection, with a 3$\sigma$ upper limit of $R<0.2$, 
	  and no line has been detected at 112~keV, as previously claimed from HEAT observations 
	  ($F_{112 \rm \, keV} < 2.4 \times 10^{-4} \, \rm ph \, cm^{-2} \, s^{-1}$).
	  The variability behaviour of \mcg\ is found to be similar to that shown by IC~4329A, with different temperatures of the electron plasma 
	  for similar flux levels of the source, while other bright Seyfert galaxies present different variability patterns at hard X-rays,
	  with spectral changes correlated to flux variations (e.g. NGC~4151).
	  }

\FullConference{8th INTEGRAL Workshop ``The Restless Gamma-ray Universe''\\
                 September 27-30 2010\\
                 Dublin Castle, Dublin, Ireland}

\begin{document}

\section{Introduction}
\begin{figure}[!b] 
\hspace{0.5cm}
\includegraphics[width=.93\textwidth,angle=0]{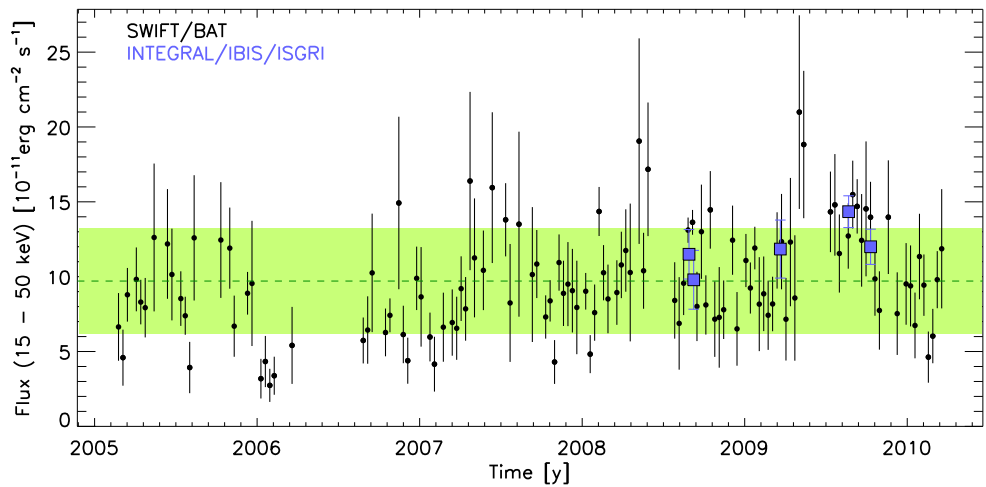} 
\caption{\mcg\ light curve in the 15--50~keV energy range as observed with \textit{SWIFT}/BAT
	 (black circles) and ISGRI (blue squares). The dashed line and the shaded area indicate
	 the average value of the hard X-ray flux measured by BAT and the dispersion from the mean.} 
\label{lc_mcg} 
\end{figure}

\vspace{-0.3cm}
The Seyfert 1.5 \mcg\ ($z = 0.0205$) is one of the brightest Seyfert galaxies in the X-rays and historically showed flux variations by a factor of 
3--4 on time scale of years (Fig.~\ref{lc_mcg}). This source has been extensively observed in the past below 10 keV, but only few observations were performed 
at hard X-rays before the advent of the present hard X-ray satellites \textit{INTEGRAL}, \textit{SWIFT} and \textit{Suzaku}.
Due to the relatively low amount of reflection and the properties of the Fe line, it has been suggested that the 
observed reflection is produced in the accretion disc and that the narrow core of the Fe K$\alpha$ line is emitted in the broad line 
region, whereas no evidences for a classical torus are found in the X-ray spectrum of this AGN \cite{bianchi10}, making \mcg\ a very peculiar Seyfert galaxy.
We present here the preliminary results of simultaneous \textit{INTEGRAL} \cite{winkler03} and \textit{SWIFT} \cite{gehrels04} pointed observations of \mcg.
Such observations aim at measuring the Comptonisation parameters for an increasing sample of Seyfert galaxies in order to test the AGN unification 
model and to provide more precise inputs for cosmic X-ray background models.
In addition, the study of the spectral variability properties of AGN can provide important information about the structure, the physics and the dynamics 
of the radiating source.
\vspace{-0.2cm}
\section{\textit{INTEGRAL} and \textit{SWIFT} observations}
\vspace{-0.3cm}
Before the 2009 pointed observation, \mcg\ had been observed by \textit{INTEGRAL} during non-dedicated observations, resulting in off-axis
angles mostly larger than $7^{\circ}$. For this reason, a relatively low effective IBIS/ISGRI \cite{lebrun03} exposure time of only 235~ks had been accumulated, 
compared to the 660~ks of field exposure.
During August and October 2009, \textit{INTEGRAL} has performed 400~ks (242 ks of effective, off-axis corrected ISGRI time) of pointed observations on 
\mcg, coordinated with 2 quasi-simultaneous \textit{SWIFT} snapshots of 2.4 and 3.6 ks, respectively. 
We have analysed all the IBIS/ISGRI data available in the archive with the source within $10^{\circ}$ from the center of the field of view, resulting
in 477 ks of effective exposure, accumulated between the beginning of the mission and October 11, 2009.
The data have been analysed using version 9.0 of the Offline Scientific Analysis Software (OSA).
The spectra have been extracted using the standard OSA spectral extraction.

The two \textit{SWIFT} observations have been performed on August 26 and October 8, 2009 and the XRT data have been analysed with 
the \textit{SWIFT} software version 3.4 distributed with the HEAsoft 6.7 package and the latest available calibration files.

\vspace{-0.1cm}
\section{Spectral analysis} 
\vspace{-0.2cm}
\subsection{Spectral slope and flux variability} 
\begin{figure}[!t] 
\hspace{2.5cm}
\includegraphics[width=.45\textwidth,angle=90]{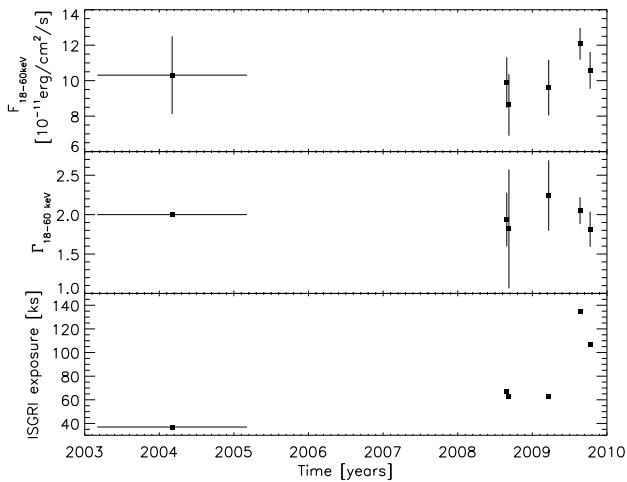} 
\caption{Evolution with time of the 18--60~keV flux (top panel) and hard X-ray photon index (middle panel)
	 as observed in ISGRI data. The bottom panel indicates the effective exposure time of the fitted
	 spectrum. Due to low statistics, the photon index of the 2003--2005 spectrum has been fixed to a value of 2.} 
\label{var_spe_mcg} 
\end{figure}

We have grouped the IBIS/ISGRI data into 6 separate periods, spanning from 40 to 170 ks of effective exposure time each, accumulated within
4 days to 2 years. In order to investigate the variations of the spectral shape and its correlation with the source flux across the mission lifetime,
we have fitted these spectra with a simple power law model. The statistic of the spectrum collected during the first period (2003--2005) does not allow to constrain 
the power law slope, which has therefore been fixed to the value $\Gamma = 2$. 
No correlation is found between the spectral slope and the 18--60 keV flux, with the flux varying by about 40\% while the slope did not significantly change 
(Fig.~\ref{var_spe_mcg}).
During the 2008--2009 \textit{INTEGRAL} observations, the source was detected with a flux larger than its average level over the last 6 years as measured
by \textit{SWIFT}/BAT\footnote{The BAT light curve in the 15--50 keV band has been downloaded from the \textit{Swift}/BAT hard X-ray transient monitoring 
pages: \emph{http://swift.gsfc.nasa.gov/docs/swift/results/transients}}
 (Fig.~\ref{lc_mcg}). 
\subsection{Comptonisation features}
\begin{figure}
\begin{minipage}[t]{17cm}
 \hspace{-0.5cm}
  \begin{minipage}[t]{7cm}
  \vspace{-6.45cm}
  \includegraphics[width=.88\textwidth,angle=-90]{fig3.ps} 
  \end{minipage}
 \hfill
 \begin{minipage}[t]{7cm}
  \hspace{-3.0cm}
  \psfig{width=.91\textwidth,angle=90,figure=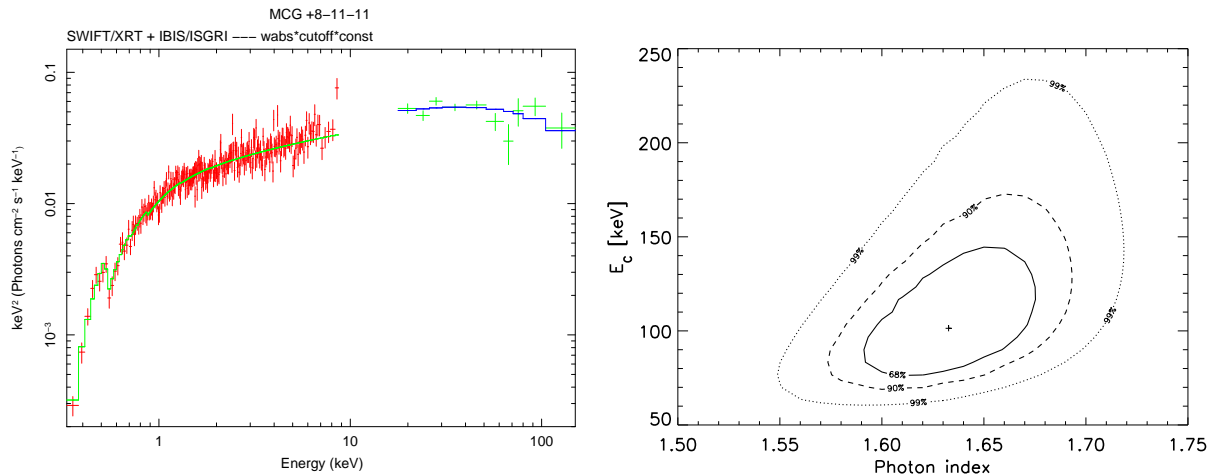}
 \end{minipage}
\end{minipage}
  \caption{\textit{Left}: Total \textit{INTEGRAL}/IBIS ISGRI spectrum (spanning the whole mission time) and \textit{SWIFT}/XRT averaged
	   spectrum of \mcg\ fitted with an absorbed cut-off power law.
	   \textit{Right}: Cut-off energy versus photon index contour plot at 68\%, 90\% and 99\% confidence level for the average XRT + total ISGRI
	   spectra.}
  \label{spe_mcg} 
\end{figure}

A more detailed analysis has been carried out on the spectra with the highest-signal-to-noise, using a more complex model in order to test the presence
of Comptonisation signatures, e.g. a high-energy cut-off and a Compton reflection hump.
Therefore, we fitted the IBIS/ISGRI spectra of August and October 2009 with the respective quasi-simultaneous 
\textit{SWIFT}/XRT data, and the total ISGRI spectrum spanning 7 years of the \textit{INTEGRAL} mission together with the average XRT spectrum.
Preliminary results indicate that while the October 2009 spectrum is best fitted with a simple absorbed power law ($\chi^2_{\rm red} = 0.93$, 174 d.o.f.), 
the August 2009 spectrum shows a prominent curvature, best modeled with the addition of a cut-off at 
$E_{\rm C}= 115{+117 \atop -43} \rm \, keV$, and resulting in a significant improvement of the reduced $\chi^2$ ($\chi^2_{\rm red} = 0.99$, 126 d.o.f., 
F-test probability = $4 \times 10^{-4}$).
Similarly, the total spectrum requires the addition of a cut-off at $E_{\rm C}= 101{+48 \atop -27} \rm \, keV$ ($\chi^2_{\rm red} = 1.0$, 265 d.o.f., 
F-test probability = $2 \times 10^{-8}$; Fig.~\ref{spe_mcg}).
A summary of the best parameters obtained fitting these three ISGRI spectra with an absorbed cut-off power law model is provided in 
Table~\ref{tab}\footnote{A cross-calibration factor has been added to all fitting models, and left free to vary. This factor has been found
to assume values of $C = (0.8-0.9) \pm 0.1$ for the ISGRI spectra with respect to the XRT ones.}.
Only a 3$\sigma$ upper limit could be derived for the amount of Compton reflection present in the source ($R < 0.2$, when applying a {\sc pexrav} model
to the total spectrum) and for the flux of the Fe K$\alpha$ line 
($F < 3.1 \times 10^{-4} \, \rm ph \, cm^{-2} \, s^{-1}$, when fixing the rest frame line energy and the line width at $E_{\rm Fe} = 6.4 \rm \, keV$ and 
$\sigma_{\rm Fe} = 0.17 \, \rm keV$, respectively; \cite{matt06}).
We tried to fit the spectra with a more physical model, i.e. {\sc compps}, but it was not possible to significantly constrain its parameters. 
A more complete and detailed analysis of \mcg\ combining \textit{INTEGRAL}, \textit{SWIFT}, \textit{XMM-Newton}, 
\textit{RXTE} and \textit{Suzaku} data and using the {\sc compps} model is ongoing \cite{lubinski11} and its preliminary 
results are in agreement with those presented here.
\subsection{The 112~keV line}
\vspace{-0.1cm}
An emission line at 112~keV was detected in the spectrum of \mcg\ during a balloon flight of the HEAT telescope in July 1995 with a flux of 
$9 \times 10^{-4} \rm \, ph \, cm^{-2} \, s^{-1}$ \cite{perotti97}.
As simultaneously the veto system of the telescope registered a strong gamma-ray flux, Perotti et al. have suggested
that the 112~keV line could be due to double-Compton scattering of a 511~keV line source incident on an external cloud.
Later \textit{CGRO}/OSSE observations did not detect the line with an upper limit flux of $0.9 \times 10^{-4} \rm \, ph \, cm^{-2} \, s^{-1}$ 
\cite{grandi98} implying that the line, if present, is variable.
In the ISGRI data presented here, we do not detect any line at 112~keV. We provide a 3$\sigma$ upper limit of $2.4 \times 10^{-4} \, \rm ph \, cm^{-2} \, s^{-1}$ 
for the line flux, when fixing the position and the width of the line to $E_{\rm line} = 112 \rm \, keV$ and 
$\sigma_{\rm line} = 32 \, \rm keV$ \cite{perotti97}.
This upper limit is 4 times lower than the flux reported by Perotti et al. and consistent with the upper limit found with OSSE.
The source was not detected by SPI and, in any case, due to its sensitivity and spectral resolution, ISGRI is a better instrument than SPI to 
detect a rather broad line at energies below 200~keV.
\begin{table}[!b]
\begin{center}
\begin{tabular}{c c c c c c c}
\hline\hline                
ISGRI obs.  & ISGRI exp. & $N_{\rm H}$               & $\Gamma$ & $E_{\rm C}$ & $F_{2-10 \, \rm keV}$ & $F_{20-60 \, \rm keV}$  \\   
period	  & [ks]       & $[10^{22} \,\rm cm^{-2}]$ &          & [keV]       & $\rm erg \, cm^{-2} \, s^{-1}$ & $\rm erg \, cm^{-2} \, s^{-1}$  \\   
\hline                      
\noalign{\smallskip}
August 2009  & 135 & 0.21 $\pm$ 0.03  & 1.61 $\pm$ 0.08 & $115 {+117 \atop -43}$ & $8.1 \times 10^{-11}$ & $1.1 \times 10^{-10}$ \\
October 2009 & 173 & 0.18 $\pm$ 0.02  & 1.69 $\pm$ 0.06 & $\geq 320$ & $6.6 \times 10^{-11}$ & $9.6 \times 10^{-11}$ \\
All data     & 477 & 0.19 $\pm$ 0.01 & 1.63 $\pm$ 0.05 & $101 {+48 \atop -27}$ & $7.2 \times 10^{-11}$ & $9.4 \times 10^{-11}$ \\
\noalign{\smallskip}
\hline                         
\end{tabular}
\end{center}
\caption{Best fit parameters for the XRT+ISGRI spectra of \mcg.}
\label{tab}
\end{table}
\vspace{-0.4cm}
\section{Discussion and conclusions}
\vspace{-0.3cm}
With a black hole mass of $M_{\rm BH} = (1.5-13) \times 10^7 \, \rm M_{\odot}$ \cite{bian03,middleton08}, an X-ray luminosity 
of $L_{\rm 0.1-300 \, keV} $ $= 2.7 \times 10^{44} \, \rm erg \, s^{-1}$ and assuming a twice as large bolometric luminosity, MCG~$+$8$-$11 $-$11 shows an Eddington 
ratio of $L_{\rm Bol}/L_{\rm Edd} = 0.03-0.3$, placing it among the most efficient hard X-ray selected Seyfert galaxies \cite{beckmann09}. 

Some of the past studies of \mcg\ reported the presence of a strong reflection component $R = 1-2.5$ \cite{matt06,grandi98,perola00,molina09}.
Yet, a good statistical constrain was only obtained with \textit{XMM-Newton} data but limited at energies below 10~keV \cite{matt06} or combined with
low effective exposure ISGRI data (\cite{molina09}, corresponding to the first point in Fig.~\ref{var_spe_mcg}), until \textit{Suzaku} together 
with \textit{SWIFT}/BAT observations provided a high signal-to-noise, broad band coverage between 0.6 and 150~keV \cite{bianchi10}.
The latter work finds that when the broad band spectrum is fitted, the reflection fraction drops to values of $R = 0.2-0.3$, marginally 
consistent with the upper limit we find in \textit{INTEGRAL} data.
 
During observations at hard X-rays with \textit{CGRO}/OSSE \cite{grandi98}, \textit{BeppoSAX} \cite{perola00}, 
\textit{Suzaku} + \textit{Swift}/BAT \cite{bianchi10} and in early \textit{INTEGRAL}/ISGRI data \cite{molina09}, \mcg\ showed a somewhat steeper spectrum 
($\Gamma = 1.7-1.9$) and a cut-off at higher energies ($E_{\rm C} = 130-270 \, \rm keV)$, but still consistent within the uncertainties with the value found here for the August 2009 spectrum.
A cut-off energy at $\sim 100 \rm \, keV$ corresponds to an electron plasma temperature of $kT_{\rm e} = 30-50 \rm \, keV$, towards the lower end of typical
values found in Seyfert galaxies \cite{lubinski10}.
\textit{INTEGRAL} data seem to indicate that the position of the high-energy cut-off varies on month time scale, while the spectral slope and the flux 
did not significantly change during the same period. Yet, it is important to keep in mind the uncertainties of the cut-off determination, due to the fact 
that the source is detected up to only $\sim$120~keV in these ISGRI data.

Past studies suggest that the hard X-ray emission of Seyfert galaxies is indeed observed to vary, and 
the variations can be complex and very different from object to object.
For instance, in some cases no spectral variations are observed to accompany different flux levels \cite{guainazzi99}
whereas in some other cases variations of the spectral slope are correlated with the flux \cite{papadakis09}, suggesting 
the superposition of a constant reflection component and a variable (in flux and shape) power law continuum.
On the other hand, NGC~4151 presents no changes in the slope of the underlying continuum for different X-ray fluxes but a
significant variation of the temperature of the Comptonising plasma is observed \cite{lubinski10}.
For other objects, a clear variability pattern has not been recognized yet and equivalent flux levels
can present significantly different spectral slopes (NGC~4388 \cite{fedorova10}, NGC~7172 \cite{guainazzi98}) 
or different cut-off energies (this work and IC~4329A \cite{soldi09}).

The large amount of \textit{INTEGRAL} data currently available will enable us to extend this kind of studies 
to about 20 Seyfert galaxies, among which NGC~7172, NGC~4593, NGC~2110, Mrk~509 \cite{paltani10}, as already done for other \textit{INTEGRAL} AGN 
(e.g. \cite{lubinski10,fedorova10,beckmann08,beckmann11,soldi05,beckmann11b}).

\vspace{-0.4cm}
\section*{Acknowledgments}
\vspace{-0.3cm}
\small{ 
S.S. acknowledges the Centre National d'Etudes Spatiales (CNES) for financial support.
Part of the present work is based on observations with \textit{INTEGRAL}, an ESA project with
instruments and science data centre funded by ESA member states with the participation of Russia and the USA.
We thank Peter Kretschmar for providing the script to draw the contour plot.
}

\end{document}